\documentclass{aa}
\usepackage{amsmath}
\usepackage{psfig}
\usepackage{natbib}
\usepackage{journals}
\newcommand{\figwidth}{0.5\textwidth}
\def\gx{GX~339$-$4\;}
\def\be{\begin{equation}}
\def\ee{\end{equation}}
\begin{document}

\title{Exploring the Role of Jets in the Radio/X-ray Correlations of GX~339$-$4}

\author{Sera Markoff\inst{
1}\fnmsep\thanks{Humboldt Research Fellow.  Current address:
Massachusetts Institute of Technology, Center for Space Research,
Rm. NE80-6035, Cambridge, MA 02139, USA}
\and Michael Nowak\inst{2} \and St\'ephane Corbel\inst{3} \and  Rob Fender\inst{4} \and Heino Falcke\inst{1} }

\institute{Max-Planck-Institut f\"ur Radioastronomie, Auf dem H\"ugel
69, D-53121 Bonn,
Germany 
\and Massachusetts Institute of Technology, Center for Space Research
Rm. NE80-6077,
77 Massachusetts Ave., Cambridge, MA 02139, USA
\and Universit\'e Paris VII and Service d'Astrophysique, CEA,
CE-Saclay, 91191 Gif sur Yvette, France
\and Astronomical Institute `Anton Pannekoek' and Center for High
Energy Astrophysics, University of Amsterdam, 
Kruislaan 403, 1098 SJ Amsterdam, The Netherlands}

\titlerunning{The Role of Jets in GX~339$-$4}
\authorrunning{Markoff et al.}
\offprints{sera@space.mit.edu}

\date{Received ; Accepted}

\abstract{The Galactic black hole candidate X-ray binary \gx spends
most of its time in the low/hard state, making it an ideal candidate
for modeling the assumedly low accretion phase.  The radio emission
correlates very tightly with the X-rays over more than two orders of
magnitude in X-ray flux density, suggesting that the jet plasma also
plays a role at the higher frequencies.  We compare the predictions of
our jet model, with and without acceleration, to thirteen broadband
simultaneous or quasi-simultaneous spectra over this changing flux
history.  In addition, we consider a simple standard thin disk which
transitions to an optically thin accretion flow, in order to account
for the assumedly thermal optical data seen in some observations.  A
solution without acceleration cannot describe the data without
unrealistic energy requirements, nor explain the non-thermal radio
spectrum seen during recent radio outbursts.  But because of the low
disk luminosity, and possibly the assumed disk geometry, acceleration
in the jet is limited only by synchrotron cooling and can extend
easily into the X-rays.  We present a model which can account for all
the broadband spectra included here, by changing only two parameters
in the jet model: the input power and the location of the first
acceleration zone.  However, the model is most sensitive to
changes in the jet power, the varying of which can also account for
the slope of the observed radio/X-ray correlation analytically.  At
the highest low/hard state luminosities, the synchrotron self-Compton
emission from the jet could be detectable with missions such as {\em
GLAST}, providing a way to test the extent of the synchrotron
contribution.  We conclude that jet synchrotron is a possible way to
explain the broadband features and this correlation, and discuss ways
of incorporating this component into the ``standard'' corona picture.
\keywords{X-rays: binaries -- X-rays: individual: GX~339$-$4 --
radiation mechanisms: non-thermal -- stars: winds, outflows --black
hole physics -- accretion, accretion disks} } \maketitle

\section{Introduction}

The Galactic black hole candidate (BHC) X-ray binary (XRB) \gx
displays a wide variety of spectral states, and has recently
resurfaced at a high flux level after being in an extended ``off''
state since 1999 (see, e.g., \citealt{Kongetal2000}).  It has been
extensively, and often simultaneously, observed across broad energy
bands: from the radio (\citealt{Hannikainenetal1998,Fenderetal1999b};
\citealt{Corbeletal2000}---hereafter C00; Corbel et al. 2002, to be
submitted---hereafter C02), infrared (IR) and optical
(\citealt{SoriaWuJohnston1999,ShahbazFenderCharles2001};
\citealt{CorbelFender2002}---hereafter CF02), to the X-rays
(\citealt{Wilmsetal1999}---hereafter W99;
\citealt{NowakWilmsDove2002}---hereafter NWD). These data sets are
relatively untainted by contamination from the companion star.  This,
and the fact that \gx often transitions between the canonical XRB
states, makes it an ideal source for comparing the emission processes
involved in these various states.

\gx spends the great majority of its time in the low/hard state (LHS)
(for a review of BHC states, see \citealt{Nowak1995}), at flux levels
varying over roughly three orders of magnitude in the X-ray waveband
(NWD; C02).  Its LHS spectra are fairly ``typical'' in that they show
a flat-to-inverted radio component, a weak thermal contribution, and a
hard power-law in the soft to hard X-rays
(\citealt{Zdziarskietal1998,Fender2001a,RevnivtsevGilfanovChurazov2001};
NWD).  Although no jets have yet been resolved, the radio emission
shows 2\% linear polarization, and the brightness temperature (defined
as $S_\nu=\Omega B_\nu(T_{\rm B})$ where $\Omega$ is the solid angle
of the emitting region) requires a spatial extension greater than the
binary separation (W99; C00).  At the time of writing, GX 339-4
briefly entered the very high state (VHS) and made a transient-like
radio outburst which was five times brighter than the brightest radio
emission ever observed in the LHS.  The optically thin spectrum had a
spectral index $\alpha \sim -0.6$ ($F_\nu\propto \nu^\alpha$)
(R. Fender, S. Corbel, private comm.), which is characteristic of a
transient bright jet.  This makes \gx the first source to show its
radio properties clearly in the LHS, the high/soft state (HSS) and
transient/VHS.  Combined with the fact that jets have already been
imaged in the other three persistent Galactic BHC which share similar
LHS spectra (e.g., 1E1740.7-2942, \citealt{Mirabeletal1992};
GRS~1758-258, \citealt{RodriguezMirabelMarti1992}; and Cygnus X-1,
\citealt{Stirlingetal2001}), the case for a jet in this source is now
very strong.

A special feature of \gx is that its radio emission has been observed
to correlate tightly with the X-ray emission over the entire range of
LHS luminosities, and also down to a very weak level of emission in
the ``off'' state (\citealt{Hannikainenetal1998}; C00; C02).
Similarly, during changeover to the HSS, the
seemingly linked radio and hard (20-100 keV) X-ray emission show an
anti-correlation with the assumed thermal disk
emission in the softer (2-12 keV) X-ray band (\citealt{Fenderetal1999b};
C00).  This suggests not only that the plasma leading to the
synchrotron radiation plays some role at higher energy, but also that
the jet is either quenched \citep{Fenderetal1999b} or otherwise unable
to form when the thermal disk component dominates
\citep{Meier2001,MerloniFabian2002}.

On the other hand, the LHS X-ray spectra from \gx show additional
features which have been interpreted as signs of reflection off colder
disk material, and can be fit by one of several disk-corona models as
discussed in detail in NWD.  In this paper, the authors find
correlations between the reflection covering fraction and both the
soft X-ray flux (see also
\citealt{ZdziarskiLubinskiSmith1999,RevnivtsevGilfanovChurazov2001}),
and the time lags between the soft and hard X-ray variability.
Similar to what \citet{Pottschmidtetal2000} report for Cyg X-1---which
also shows radio/X-ray correlations \citep{Brocksoppetal1999}---NWD
find that these time lags anti-correlate with the coronal compactness, and peak shortly after the transition from the HSS to the
LHS, when the assumed jet emission begins to rise.  They suggest an
interpretation that the time lags may be associated with the extended
synchrotron-emitting plasma.

While the model fits presented in NWD can well explain the X-ray
features, and suggest a link between the corona and jet, they do not
explore the nature of this link or how this emission is related to
that of the radio.  One model which can explain the tight radio/X-ray
correlations proposes that the jets directly contribute to the LHS
X-ray spectrum via synchrotron emission
(\citealt{MarkoffFalckeFender2001}---hereafter MFF), as occurs in
several AGN (e.g., 3C 66B, \citealt{Hardcastleetal2001} and M87,
\citealt{Marshalletal2002,WilsonYang2002}).  However, this model has
only been applied until now to one source, XTE~J1118$+$480, which may
be unusual in that it has extremely weak or absent reflection features
\citep{Milleretal2001}.

In this paper, we discuss the application of a jet model to the
collected simultaneous or quasi-simultaneous data sets for \gx in
order to investigate whether it succeeds in explaining the observed
radio/X-ray correlations.  We discuss how the model fares against the
broadband data and compare the model parameters for the various data
sets to each other, as well as to XTE~J1118$+$480.

\section{Data}

The 1981 data set was compiled by CF02, where it is explained in
further detail.  The X-rays come from the Ariel 6 instrument
\citep{Ricketts1983}, with simultaneous optical/IR from
\citet{Pedersen1981} and \citet{Motchetal1981}.  

All radio observations were performed with the Australia
Telescope Compact Array (ATCA) and, with the exception of the 2000
September 15 observation (to appear in C02), have been presented
previously by C00.  All observations from C00 in which the radio flux
density was weak (i.e., $\la1$ mJy) were reanalyzed for this paper.

The X-ray data presented here have been discussed previously by NWD
and C02.  All data were obtained by the Rossi X-ray Timing Explorer
(RXTE), with the exception of the 2000 September 15 data, which were
obtained by BeppoSAX.

In this work we are attempting to describe the \emph{broadband
features} of the jet model, and to simultaneously explain radio, IR,
optical, and X-ray data (taken with a wide variety of instruments).
Furthermore, we are not yet attempting to explain detailed features
within the X-ray, e.g., Fe lines or reflection features.  Therefore,
rather than forward fold the jet models through the detector response
matrices, we instead compare the models to ``unfolded'' count flux
rates.  This is not strictly fitting, as we cannot obtain chi-squared
estimates from unfolded data in this way, but for simplicity we will
use the word ``fit'' to describe the comparison to the data throughout the
paper.  

To unfold the X-ray data, the RXTE data extraction, data binning,
response matrix generation, and background subtraction were carried
out in an identical manner as presented by NWD.  We then fit models
comprised of: a multi-temperature disk blackbody (e.g.,
\citealt{Mitsudaetal1984,Makishimaetal1986}) with peak temperature
fixed at 0.25\,keV, an exponentially cutoff broken power-law with
break energy at $\approx 10$\,keV, and a (potentially) broad gaussian
line with energy fixed to 6.4\,keV. (This model is essentially
identical to that discussed by W99).  These model fits were then
unfolded through the detector response matrix and multiplied by the
ratio of the count rate data to the fitted model folded through the
detector response.  For the RXTE observations, data was considered
only for the 3-200\,keV range or less for the faintest data sets.
Furthermore, HEXTE data was normalized to the PCA data (e.g., see
W99).

\begin{table}[t]\label{tab1}
\caption{Explanation of Labels in Figures}
\begin{tabular}{cll}
\hline
\hline
{\bf Label} & {\bf Data Set$^a$} & {\bf Date (y.m.d)}\\
\hline
$97_1$ & 20181-01-01-00 & 1997.02.03\\
$97_2$ & 20181-01-02-00 & 1997.02.10\\
$97_3$ & 20181-01-03-00 & 1997.02.17\\
$99_1$ & 40108-02-01-00 & 1999.04.02\\
$99_2$ & 40108-02-02-00 & 1999.04.22\\
$99_3$ & 40108-02-03-00 & 1999.05.14\\
$99_4$ & 40105-02-02 & 1999.06.25\\
$99_5$ & 40108-03-01-00 & 1999.07.07\\
$99_6$ & 40108-03-02-00 & 1999.07.29\\
$99_7$ & 40108-03-03-00 & 1999.08.17\\
$99_8$ & 40104-01-08 & 1999.08.28$^b$\\
$00$ & 21136001 & 2000.09.15$^c$\\
\hline
\end{tabular}

\small $^a$Data from W99, C00, C02 and NWD. All but 00 (which is BeppoSAX) are labeled
according to RXTE conventions.\\
$^b$Date for X-ray observation.  Radio taken on 1999.09.01.\\
$^c$Optical data shown in Fig.~\ref{fig4b}b were taken 3 months prior.
\end{table}

As discussed by C02, simultaneous RXTE/ASCA observations indicate a
probable faint background source that contaminates the RXTE spectrum---
but not the narrow field of view ASCA spectrum---at \emph{extremely}
low flux levels.  The RXTE observation of 1999 July 29 is assumed to be
heavily dominated by this contaminating source, and this spectrum,
multiplied by 0.78, is subtracted, before model fitting and data
unfolding, from all RXTE observations occurring later than the
observation of 1999 May 14.  This normalization was determined by
comparison of spectra obtained simultaneously by RXTE and ASCA on two
separate occasions (the required normalization constants for these two
observations were 0.73 and 0.83; see C02). Note that the 2000
September 15 BeppoSAX observation is most likely \emph{not} contaminated by
this background source, which could easily be half a degree away and
still affect the RXTE observations, as BeppoSAX has a narrower field
of view than RXTE.  A more detailed discussion of this background
source appears in C02.  

The dates of the observations are listed in Table 1.  Data sets
labeled 20181 were first discussed by W99.  Data sets 40108-02 were
first discussed by NWD.  Data sets 40108-03, first appearing in C02,
were part of the same series of observations originally discussed by
NWD; however, these latter observations were too faint for
simultaneous spectral and X-ray variability analyses.  Observations
40105-02-02, 40104-01-08 and 21136001 were part of a series of
observations to study \gx\ in quiescence, and are discussed by C02.

\section{Model}

The details of the jet model can be found in MFF, and references
therein, and we give a brief summary here.  We consider that an
accretion disk may contribute to the near infrared (NIR), optical and
possibly soft X-ray emission, but that the radio through at least far
infrared (FIR) are due to synchrotron emission from the jets, which
could possibly extend through the NIR to the X-rays.  The inverse
Compton (IC) upscattering by the hot jet electrons of both the thermal
disk photons, as well as the synchrotron photons, is also included.

Our disk model is relatively simplistic, since we are currently
exploring its limits in terms of contribution and there are already
many detailed disk models in existence (see \citealt{Poutanen1998}; NWD
for reviews).  We here consider one commonly invoked model in which a
standard thin, optically thick disk (SD) \citep{ShakuraSunyaev1973}
exists down to some transition radius $r_{\rm tr}$, at which point the
flow becomes hotter, optically thin and relatively non-radiative
(e.g., \citealt{EsinMcClintockNarayan1997}).  This type of
non-radiative inner disk has also been invoked for low-luminosity AGN
such as NGC~4258 \citep{Yuanetal2002}.  The thermal bump---seen in
several of the \gx data sets in the optical band---is assumed to
originate in the disk, which also provides seed photons for jet IC as
mentioned above.  We assume for now that the jet is oriented
90$^\circ$ from the plane of the disk, but in the future we will
consider other orientations (see, e.g., \citealt{Maccarone2002}).

In their study of the BHC XTE~J1118+480, MFF interpreted the
extreme-ultraviolet (EUV) upper limits \citep{Hynesetal2000} as the
hottest, and thus innermost, emission from this outer disk. Therefore,
they could place limits on its maximum temperature and luminosity.
They derived a location for the transition radius of $r_{\rm tr} \sim
10^2-10^3 r_g$ which is consistent with other models (e.g.,
\citealt{Liuetal1999,Esinetal2001}).  However, no EUV data exist for
\gx, and so we constrain $r_{\rm tr}$ with the low-frequency X-ray
data.  We have also made the outer disk slightly more realistic than
the simple one-temperature component we used before by considering a
multi-temperature blackbody spectrum as its representation (e.g.,
\citealt{ShakuraSunyaev1973,Mitsudaetal1984,Makishimaetal1986}).  We
then explore its possible geometry using the data sets where its
signature seems the most prevalent.

A significant problem for modeling \gx is that most of its system
parameters are not well constrained.  The only report of an orbital
period based on optical photometry is from \citet{Callananetal1992},
who found 14.8 hours.  This value remains unconfirmed (though
\citealt{Cowleyetal2002} reconsidered the problem and found a similarly
tentative period of 16.8 hours).  However, it has been adopted over
the years to limit the mass function, which is also problematic since
the companion star is still unclassified.  Finally, because the system
has not yet been resolved, the inclination angle is also unknown but
limited to $\la 60^\circ$ because of the lack of eclipsing.  Based on
reasonable assumptions, a distance of 4 kpc has been adopted (see
\citealt{Zdziarskietal1998}; W99; but also
\citealt{ShahbazFenderCharles2001}, who argue for $\ge 5.6$ kpc) and a
mass for the black hole ranging from $\sim 3-6 M_\odot$, where we here
use $5 M_\odot$.

We fix the inner temperature of the SD, $T_{\rm in}$, based on the
1981 data set.  We then estimate the outer radius, $r_{\rm out}$, of
the SD based on the physical parameters above, the best fit
inclination angle, $\theta_i$, and assuming a companion of $\sim0.4
M_\odot$, consistent with the 16.8 hour period.  The exact luminosity
contributed by the SD, $L_{\rm d}$, is not uniquely determined by the
data, and so we can only roughly  estimate the accretion rate, $\dot M_{\rm
d}$.  Because of the simplicity of our disk model, this is not a main
focus of this work, but we want to be sure we are consistent in terms
of the energy budget.  We express $L_{\rm d}$ in units of $L_{\rm
Edd}=1.25\times10^{38}\left(\frac{M_{\rm BH}}{M_\odot}\right)$ erg
s$^{-1}$.

For accreting black holes it has been argued that the jet power is of
order $Q_{\rm j}\sim q_{\rm j}\dot M_{\rm d} c^2$ with an efficiency
inferred to be of order $q_{\rm j}=10^{-3}-10^{-1}$
\citep{FalckeBiermann1999,Meier2001}.  We consider $Q_{\rm j}$ a free
parameter in our model, also expressed in units of $L_{\rm Edd}$ erg
s$^{-1}$, and which we check for consistency with our limits on $\dot
M_{\rm d}$.  While the jet formation itself is still a looming
question in the field, the physics of calculating most of the jet
emission is relatively straightforward because the flat-to-inverted
spectrum stems from the part of the jet where it is basically
undergoing free expansion.  We briefly summarize the jet component as
follows (and see also \citealt{FalckeMarkoff2000}; MFF).

At the inner edge of the optically thin accretion flow, hot plasma is
ejected out from symmetric nozzles, where it becomes supersonic.  
We assume free jets which accelerate along their axes only due to
their longitudinal pressure gradients.  This is the simplest scenario
and provides a lower limit to the final velocity distribution.  The
velocity field along the jet is thus uniquely determined from the
Euler equation (see, e.g., \citealt{Falcke1996}), and at large
distances $z$ from the base has a dependence $\gamma_{\rm
j}(z)\beta_{\rm j}(z) \propto \sqrt{\ln(z)}$.  Thus, the maximum value
of the bulk Lorentz factor $\gamma_{\rm j}$ is roughly determined by
the distance at which the lowest frequency radio emission is emitted
and is dependent on other parameters in this model but is itself not a
fitted parameter.  For both \gx and XTE~J1118+480, we find
$\gamma_{\rm j, max}\sim 3$, which is consistent with the low beaming
factors suggested by the recent work of \citet{GalloFenderPooley2002}.

We further assume the jets expand sideways with their initial proper sound
speed, $\gamma_{\rm s}\beta_{\rm s}c\simeq0.4c$ for a hot
electron/cold proton plasma.  Beyond the nozzle region, the radius
$r(z)$ is related to the distance $z$ by the Mach number along the
jet, $\gamma_{\rm j}\beta_{\rm j}/\gamma_{\rm s}\beta_{\rm s}$.

We choose the simplifying assumption that the internal energy,
dominated by the magnetic field, is equal to the bulk kinetic energy
of the particles, consistent with a magnetic launching mechanism.  The
plasma is assumed to originate in the hot accretion flow and therefore
contains equal numbers of protons and electrons, with the exact
temperature of the electrons at the base of the jet, $T_e$, remaining
a fitted parameter.  A process such as pair-loading via interactions
with the disk photon field will not be efficient enough in the LHS to
be significant (see, e.g., \citealt{SikoraMadejski2000}).  Under these
assumptions, the magnetic energy density and particle density in each
jet can thus be defined as $B^2(z)/8\pi\simeq0.25Q_{\rm
j}/(c\gamma_{\rm j}\beta_{\rm j} \pi r(z)^{2})$ and
$n^2(z)\simeq0.25Q_{\rm j}/(c\gamma_{\rm j}\beta{\rm j} m_p c^2 \pi
r(z)^{2})$.  Taking into account the non-constant velocity field, the
dependencies of the magnetic field, $B$, and density, $n$, on distance
are then similar to, but slightly stronger than, the canonical
$z^{-1}$ and $z^{-2}$ dependencies for conical jets, respectively
\citep{BlandfordKoenigl1979,HjellmingJohnston1988,FalckeBiermann1995}.
In this way, the basic physical properties governing the emission at
each point in the jet are fixed after specifying the jet power $Q_{\rm
j}$ and nozzle scale $r_0$, which determine $B_0$ and $n_0$.

The thermal particles travel along the jet until they encounter an
acceleration region which begins at $z_{\rm sh}$.  An acceleration
process is inferred because of the observed optically thin synchrotron
power-law seen in the recent transient radio events, indicating
the presence of non-thermal particles.  This is a standard process
invoked to explain many features of AGN jet emission
\citep{MarscherGear1985}.  We consider the process to be diffusive
shock acceleration, but it could really be one of a few possible
scenarios (e.g., stochastic acceleration) which lead to a non-thermal
power-law with spectral index $p$ ($\frac{dN}{dE}\propto E^{-p}$).
For standard shock acceleration theory, $p\sim2.0-2.5$ (e.g.,
\citealt{JonesEllison1991}).  In order to maintain the non-thermal
particles out along the jet to account for the entire radio spectrum,
we assume distributed acceleration after $z_{\rm sh}$.  This is because
the cooling timescales after acceleration are very fast compared to
the dynamical timescales, a problem which has also required the
assumption of distributed acceleration in the two best studied AGN
jets, M87 and 3C 273 (\citealt{Meisenheimeretal1996,Jesteretal2001},
but see also \citealt{Perlmanetal2001}).

We assume that the accelerated distribution extends from the peak of
the thermal distribution, with Lorentz factor $\gamma_{\rm
e}\sim4\cdot T_{\rm e,10}$.  As discussed in MFF, the maximum
$\gamma_{\rm e}$ occurs when the acceleration rate (assuming Bohm
diffusion) is matched by the cooling rate giving
\begin{equation}
\gamma_{\rm e,max}\sim10^8\,\left(\xi
B\right)^{-0.5}\left(\frac{u_{\rm sh}}{c}\right),
\end{equation}
where $u_{\rm sh}$ is the shock speed in the plasma frame and $B$ is
the magnetic field in Gauss.  The
parameter $\xi$ is the ratio between the diffusive scattering mean
free path and the gyroradius of the particle.  It measures how many
times the particle gyrates per shock crossing, where the most
efficient acceleration possible is when it crosses the shock once per
gyroradius, thus giving a lower limit of $\xi=1$.  For
quasi-perpendicular shocks $\xi \le c \gamma_e\beta_e/u_{\rm sh}$,
although in the case of a jet the field is likely quite tangled in
which case an upper bound on $\xi$ is not known.  While we do not know
the exact value of $u_{\rm sh}$, it must be larger than the speed of
sound at the shock, $c_{\rm s}$, and no higher than the local bulk
velocity.  For very relativistic particles $\xi$ can be quite
large, but conservative values lie in the $\sim 10-100$ range (e.g.,
\citealt{Jokipii1987}).  If we define as a reference value $\xi=\xi_2
100$, the maximum synchrotron frequency is
\begin{equation}
\nu_{\rm max}=0.29 \nu_{\rm c} \simeq 1.2\cdot 10^{20} \xi_2^{-1}
\left(\frac{u_{\rm sh}}{c}\right)^2 \;\;\; {\rm Hz}
\end{equation}
where $\nu_{\rm c}\simeq\frac{3}{4\pi} \gamma_{\rm e,max}^2
(eB)/(m_{\rm e} c)$ is the critical synchrotron frequency. This
maximum corresponds approximately to the rollover of the power-law
cutoff and in order to explain the ``canonical'' 100 keV cutoff for
$\xi\approx100$, we need $u_{\rm sh}\ga c_{\rm s}$, consistent with our
physical expectations.  This cutoff is not dependent on the magnetic
field, the jet power, or the shock location as long as we are in the
synchrotron cooling dominated regime.  Because we would expect XRBs to
have similar shock structures, we should get roughly similar cutoffs
for different sources and accretion rates.  In this sense $\xi$ and
$u_{\rm sh}$ provide a reasonable physical explanation for the
location of the cutoff in much the same way the coronal energy
fraction and temperature determine it in coronal IC models (e.g.,
\citealt{SunyaevTruemper1979}).

The location of the initial acceleration region is determined by the
frequency where the flat, highly self-absorbed synchrotron spectrum
turns over into the optically thin power-law produced at the shock.
In MFF, we showed that from back-extrapolating the X-ray power-law,
this maximum self-absorption frequency should occur somewhere in the
IR/optical regime at $\sim 10^{14}-10^{15}$ Hz for XTE~J1118+480.  It
is interesting to note that this now holds true for several other BHC
sources in the LHS including \gx and V404 Cyg (CF02; Brocksopp et
al. 2002, in prep., \citealt{GalloFenderPooley2002}, Gallo et al., MNRAS, in prep.). 

There are thus seven free parameters relating to the jet component,
$Q_{\rm j}$, $T_e$, $z_{\rm sh}$, $r_0$, $p$ and $\theta_i$.  These
are not, however, entirely independent of each other. Furthermore, in
our philosophy of comparing broadband spectral features to the jet
model, we can choose reasonable constraints for a number of these
parameters.  For instance, if the X-rays are due to synchrotron
emission, the spectral index of the accelerated particles is uniquely
determined ($\alpha_X=\frac{(p-1)}{2}$).  Starting with reasonable
assumptions of the underlying disk/jet geometry in order to constrain
$T_e$ and $r_0$, $Q_{\rm j}$ and $\theta_i$ can then be determined
from the radio spectrum because they fix its normalization and
spectral index, respectively.  The beginning of the acceleration
region, $z_{\rm sh}$, is then uniquely determined by where the
turnover must occur to link the X-ray to the radio spectrum.  For the
purposes of the spectral models that we describe here, the critical
fitted parameters are $Q_{\rm j}$ and $\theta_i$, as they most
directly determine the radio spectrum.  For $\theta_i$, the only
observational constraint is that it is $\la 60^\circ$.  

As described above, there are also additional parameters describing
the cutoff in the X-rays as well as the disk component, and we will
also argue below for an additional irradiation component.  However,
these are not an integral part of our model, in that they do not
constrain the conditions of the radiating plasma in the jet.  We
include them for completeness, and to open the way for more complex,
combined disk/jet models.

Some fraction of the emitted X-rays from the jet base will either
directly interact with, or be scattered by hot electrons into, the
cold disk, resulting in reflection.  This calculation is not attempted
here, however, since again this is very dependent on the disk geometry
and we are focusing on the role of the jet.  This, as well as other
issues, will be the focus of an upcoming paper.

\section{Results}

\subsection{1981 data set: constraining the \gx model parameters}

\begin{figure*}[t]
\centerline{\hbox{\psfig{figure=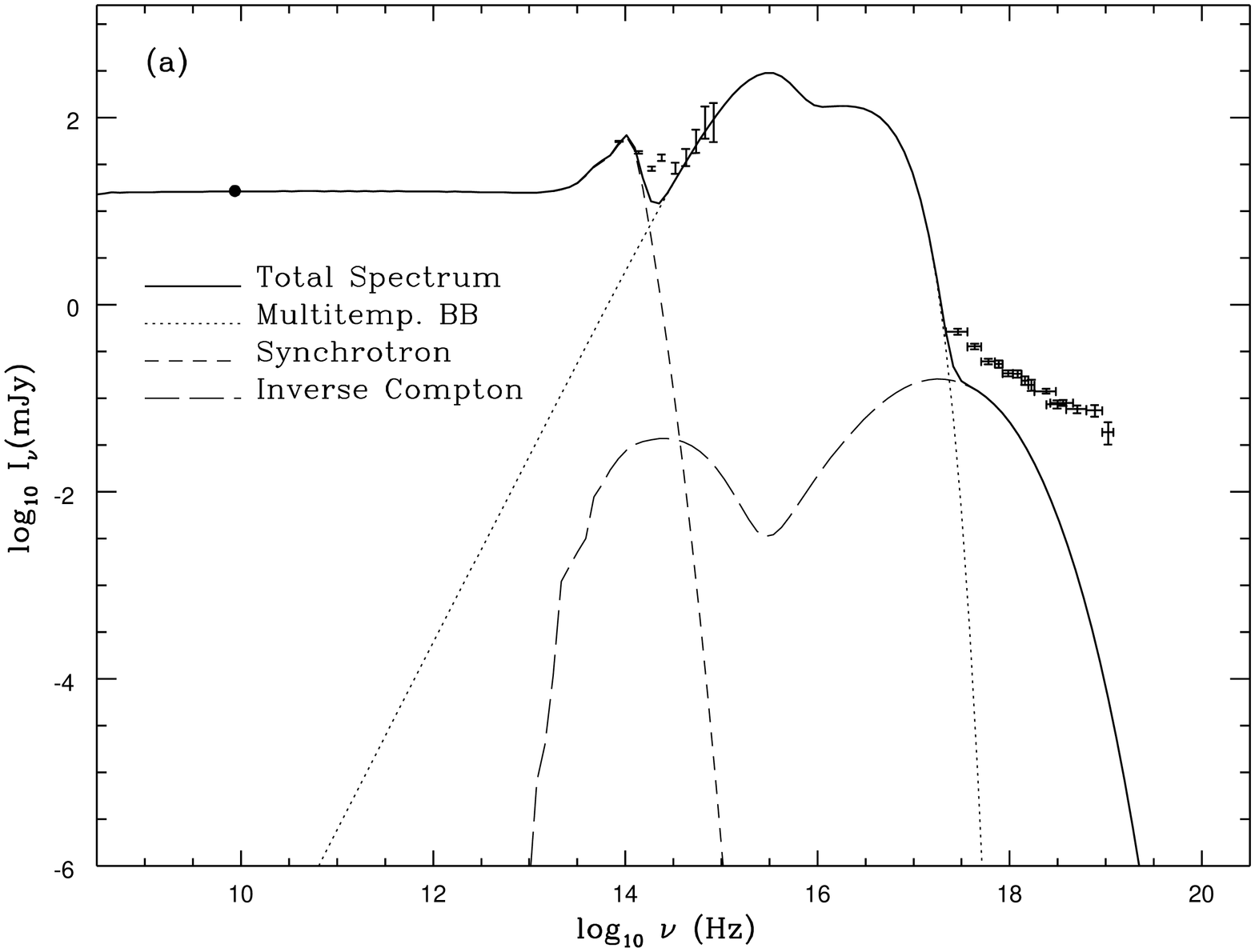,width=.5\textwidth,angle=-90}\hfill\psfig{figure=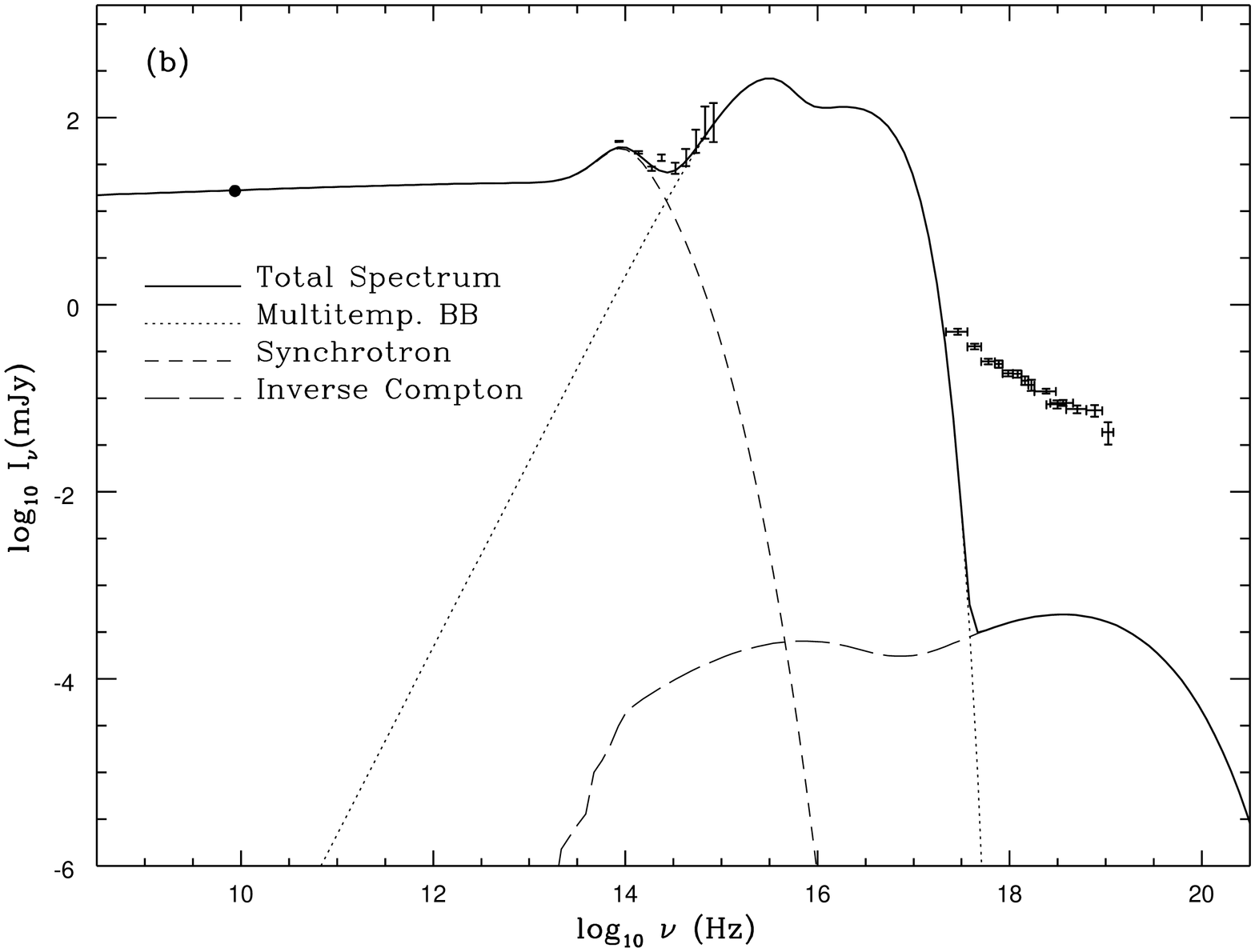,width=.5\textwidth,angle=-90}}}
\caption[]{Two fits to the \gx 1981 data sets for the case of no
particle acceleration in the jet, data references contained in CF02.
The solid thick line is the total spectrum, the dotted line is the
multi-temperature blackbody SD disk plus irradiation blackbody
contribution , the short-dashed line is the synchrotron emission and
the long-dashed line is the inverse Compton (IC) upscattered disk and
jet photons. In both panels, the IC components do not include the
outer part of the jet as its contribution is orders of magnitude under
the data.  (a) The fit for the nearly non-relativistic initial
electron temperature of $2\times10^9$ K and $r_0=5\times10^3\;r_g$,
(b) The fit for a temperature similar to the maximum derived for
another XRB source, XTE~J1118+480 (see MFF), of $2\times10^{10}$ K,
with $r_0=2\times10^3\;r_g$.}\label{fig1}
\end{figure*}

Of all the data sets for \gx considered here, the most interesting is
that of May 1981 (see Fig.~\ref{fig1}), which is at the highest flux
so far for which simultaneous broadband data have been published in
the LHS.  The IR/optical wavebands seem to indicate the clear presence
of both jet and disk components, which has not been seen explicitly in
any other XRB.  The IR band shows a first component with a negative
slope suggesting that we are seeing the expected turnover from the
optically thick to optically thin regimes.  The shape of this spectrum
at higher frequencies is hidden under a component which is likely due
to thermal emission from the SD, as indicated by the sharp rise in the
optical points.  However, if the simultaneously measured X-rays are
traced back to the IR, they line up with the turnover remarkably well,
supporting their interpretation as synchrotron emission.

\begin{figure*}[t]
\centerline{\psfig{figure=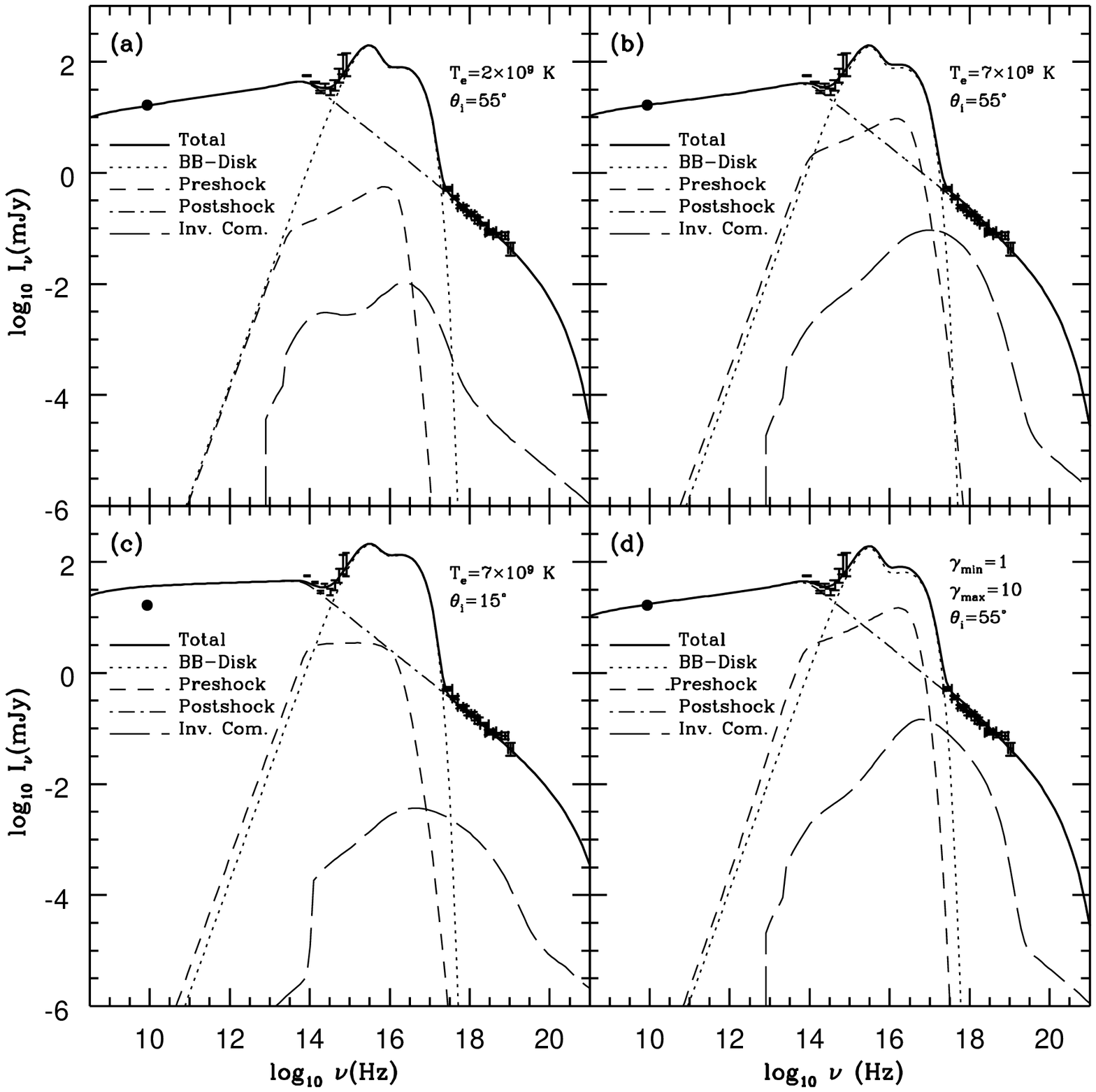,width=5in}}
\caption[]{\label{fig2}Four fits for the 1981 data set shown in
Fig.~\ref{fig1}, with the radio point extrapolated from the
correlation curves of C00 and C02.  The solid thick line is the total
spectrum, the dotted line is the multi-temperature blackbody outer disk
plus single blackbody irradiation contribution, the short-dashed is
the synchrotron emission from the jet before the shock acceleration
region, the dot-dashed line is the synchrotron emission after
acceleration, and the long-dashed line is the inverse Compton
upscattered disk and jet photons.  (a) The fit for the mildly
relativistic case of $T_{\rm e}=2\times10^9$ K, which requires an
unrealistically large jet power of $Q_{\rm j}\sim0.6 L_{\rm Edd}$.
(b) A fit with the highest temperature electrons allowed which can
give a good fit to the broadband spectrum, $T_{\rm e}=7\times10^9$.
This solution also gives a more realistic jet power of $Q_{\rm
j}\sim0.3 L_{\rm Edd}$.  (c) The case for the same temperature but
with a much smaller inclination angle, requiring only $Q_{\rm
j}\sim3\times10^{-2} L_{\rm Edd}$, but which cannot fit the
extrapolated radio point.  (d) The resulting fit from a power-law of
particles at the base of the jet, rather than a thermal distribution,
as may be expected near an accretion shock.  This solution also
requires a large jet power, $Q_{\rm j}\sim0.5 L_{\rm Edd}$, however.}
\end{figure*}

While no simultaneous radio observation exists, an estimate of the 8.6
GHz radio flux can be found by extrapolating the radio/X-ray
correlation curve to higher fluxes (C00; C02 and see Fig.~\ref{fig6}),
and is shown as a point in Figs.~\ref{fig1} \& ~\ref{fig2}).  As
mentioned above, several other LHS sources show this interesting
``turnover coincidence''.  In the compact cores of most flat-spectrum
AGN, the same kind of turnover is observed in the mm range (e.g.,
\citealt{Bloometal1994}), but the more compact scales in XRBs push this
to higher frequencies (e.g., \citealt{Falckeetal2001}).  

The clear presence of the optically thick-to-thin turnover makes the
1981 data set a strong test of any theoretical model.  Furthermore the
radio/X-ray correlation, illustrated by the full set of simultaneous
data shown here, suggest that any model for the X-rays from this
source must address the lower-frequency data as well.  This means that
if---as is usually assumed---the X-rays are due entirely to IC
upscattering off hot electrons in the base of the jet/coronal plasma,
one has to show that these same electrons can reproduce the
flat-to-inverted radio spectrum via their synchrotron emission, or
that some other very strong coupling exists between ``corona'' and
``jet''.  As \citet{BlandfordKoenigl1979} explain, to fit the
flat-to-inverted radio spectrum particularly in the cm-band, one
needs a conical geometry with outflowing plasma, otherwise the emitted
synchrotron spectrum will be too inverted.  We therefore think that a
corona which has no relation to the jet cannot explain the data for
\gx.  However, the corona may provide the launching point for the jet
\citep{MerloniFabian2002} or even comprise the base of the jet itself
\citep{Fenderetal1999b}.  Interestingly, \citet{Beloborodov1999} found
that magnetized plasma moving away from the disk with a velocity of
$\sim 0.3c$ can explain the weak reflection features in Cyg X-1, which
would be consistent with this picture.

In order to understand the IR/optical data, we must first consider how
the disk may be contributing in this region.  Unfortunately,
addressing the thermal disk component is not a uniquely defined
problem for \gx.  If we assume the SD component extends up to the
first X-ray data points and then drops off, we find $T_{\rm in}=10^6$
K, which is the same value which W99 used for their fits to this
source.  We keep $T_{\rm in}$ fixed to this value for all models
discussed in this paper.  However, a naive fit to the rising optical
data with a multi-temperature blackbody would then require almost the
entire $L_{\rm Edd}$ being channeled into the SD emission.  For the
LHS, which is defined by its assumedly low accretion rate, this seems
unrealistically high.  One way around this is to assume a lower SD
luminosity, and that the optical points are due to the irradiation of
the outer disk material by intercepted X-rays.  For the mass and
inclination angle we use here, the outer disk radius would be on the
order of $r_{\rm out}\approx10^5$ cm.  We use this as the nominal
scale for the X-ray irradiated shell, noting that there could easily
be a factor of a few leeway due to the large uncertainties in the
system's binary parameters.  To fit the optical data, the temperature
of this single blackbody needs to be  $T_{\rm BB}\sim5\times10^4$ K.  This
high temperature is not in conflict with possible disk instability models
(e.g., \citealt{CannizzoShafterWheeler1988}) because we are observing
already in outburst, after the effects of the instability would be
felt.  

For the models presented in Fig.~\ref{fig1}, we fix the SD luminosity
to a characteristic value of $L_{\rm d}\sim 0.1 L_{\rm Edd}$, and then
make up any deficit in the optical region via the irradiation
component.  The ensuing fraction of the jet power contained in the
irradiated component then can be determined due to the fact that an
IR/optical break is observed.  For the other data sets included in this
paper, however, this break is not observationally resolved.  For those
models we therefore assume that the luminosity of the irradiated
emission can be linked to the power in the jet, and we set $L_{\rm
BB}=0.1 Q_{\rm j}$ as a canonical value.

It turns out that the 1981 data set is in fact so luminous that it
creates a problem for our assumed black hole mass of $5 M_\odot$.
Even assuming $L_{\rm d}\sim 0.1 L_{\rm Edd}$ in fact requires $\dot
m\sim0.5$ (in Eddington units) for the fixed $T_{\rm in}=10^6$ K,
which is much larger than one would expect for the LHS according to some
models (e.g., \citealt{EsinMcClintockNarayan1997}).  We could solve
this problem by choosing a lower $L_{\rm d}$ and $T_{\rm in}$, thus
having the SD contribution truncate before the X-ray data.  However,
then this only increases the need for more intercepted jet power in
the irradiated component, which we already require to be at the upper
bound of physical limits.  We discuss the parameters in detail for the
models below, but feel this problem is too unconstrained at this time
to make any conclusions about the exact nature of the disk
contribution.  

\subsection{1981 data set, no acceleration}

Assuming that jets are responsible for the radio emission, the first
issue to explore is how the solution would look if the jet plasma
never encounters an acceleration region.  This would be the simplest
case model, where a portion of the thermal disk or coronal plasma is
simply advected into the outflow, retaining its thermal or
quasi-thermal distribution.  In this case, the synchrotron spectrum
from each segment of the jet will have a ``hump'' shape peaked around
a particular self-absorption frequency determined from the physical
parameters for the segment.  The combination of all these humps, each
located at a different frequency, gives a flat-to-inverted radio to at
least IR spectrum which can be seen in Fig.~\ref{fig1}.  The slope of
the spectrum depends on the amount of expansion and the velocity
profile along the jet, as well as on $\theta_i$.

As explained in the previous section, the electron temperature at the
base of the jet is a free parameter, and so we explore two extremes
for this scenario.  The first, shown in Fig.~\ref{fig1}a, is at the
lowest end of the relativistic scale with electrons entering the jet
base with initial $T_{\rm e}\sim2\times10^9$ K, and in
Fig.~\ref{fig1}b we show a more relativistic solution, with $T_{\rm
e}=2\times10^{10}$ K, the maximum temperature we could accommodate in
our modeling of XTE~J1118+480 (MFF).   In order to bring the highest
frequency ``hump'' from the part of the jet closest to the accretion
disk to low enough frequencies to come near the turnover in the IR
data, we must vary the nozzle radius $r_0$ and the input power $Q_{\rm
j}$, thus controlling the magnetic field $B_0$ and density $n_0$ at
the base of the jet while the temperature remains fixed.  Similarly we
must adjust $\theta_i$ to fit both the radio and the IR together with
synchrotron, which changes the power requirements because of the
beaming.

With the requirement that the non-accelerated models adequately
describe the IR data as well as the maximum self-absorption turnover,
the jet parameters for the fit shown in Fig.~\ref{fig1}a are then
$r_0=5\times10^3\;r_g$, $Q_{\rm j}=3.7 L_{\rm Edd}$ and
$\theta_i=8^\circ$, the former two being obviously unlikely values.
The extremely high value of the the nozzle radius is necessary to
decrease the self-absorption frequency to a value low enough to fit
the IR data, which because of the resulting larger self-absorption
also leads to a very peaked contribution from the denser nozzle.  This
does not seem feasible under the assumption that the jet originates at
the inner edges of the accretion flow, or in the corona above it, and
the match to the IR data itself is not particularly good.  The
``nozzle bump'' is due to our assumption of this element of the
geometry, as required for jet models of Sgr A* and LLAGN (e.g.,
\citealt{FalckeMarkoff2000,Yuanetal2002}).  Without a nozzle region,
one could find a solution with a somewhat larger $\theta_i$, but then
this would require an even higher $Q_{\rm j}$ because the emission
would be beamed away from the observer.  The value of $Q_{\rm j}$ here
is already several times higher than $L_{\rm Edd}$ for a 5 $M_\odot$
central object. The irradiation component can provide the required
rise in the optical data if the outer disk intercepts a small fraction
of the jet power in X-rays, $L_{\rm BB}\sim0.03 Q_{\rm j}$.  Overall,
however, this model is not so satisfactory.  

The observations are slightly better described, at least at
low-frequency, for the more relativistic case shown in Fig.~\ref{fig1}b.  The
higher temperature electrons lead to less self absorption, giving a
broader nozzle bump and requiring a somewhat smaller nozzle radius of
$r_0=1.2\times10^3\;r_g$ with a more reasonable $Q_{\rm
j}=1.4\times10^{-2}L_{\rm Edd}$ for $\theta_i=15^\circ$.  This nozzle
width is still rather untenable, however, in the context of jet/corona
models.  For this scenario we also run into more trouble with the disk
modeling, because we would require the outer disk to intercept roughly
$6 Q_{\rm j}$ in X-ray luminosity to account for the optical data.  

For both cases shown in Fig.~\ref{fig1}, the IC emission in the X-rays
is dominated by the upscattered thermal disk emission but, because of
the low densities, lies under the data and also has the wrong shape.
In this and all other figures in this paper, we do not include IC of
photons from far out in the jet, since they are orders of magnitude
under the data and will not affect the fit.  Changing the disk
geometry so that the thin disk underlies the base of the jet/corona
may provide more IC emission, but would still not address the
fundamental problem of explaining the radio thru IR data with
reasonable parameters.

In summary, the energy budget in the non-accelerated case is not large
enough to create enough synchrotron to fit the radio for a
non-relativistic plasma, since the required $Q_{\rm j}$ for the
already weakly relativistic temperature of Fig.~\ref{fig1}a is
super-Eddington.  Considering that the radio through IR bands are
likely due to jet synchrotron in the LHS, we conclude, based on all
the above difficulties, that a model which does not include particle
acceleration in the jet is unlikely.  The extent to which the energy
budget falls short is in part due to assumptions of the equipartition
of energy, as described earlier.  However, corona or other accretion
flow models which attempt to explain the low-frequency data without a
standard jet solution do not seem very feasible.

\begin{figure}[t]
\centerline{\psfig{figure=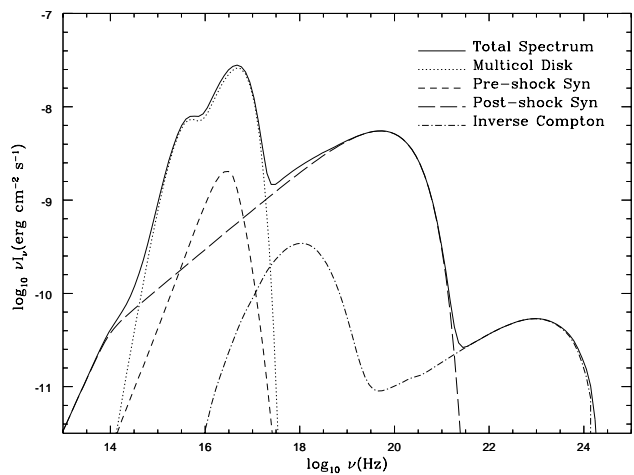,angle=-90,width=\figwidth}}
\caption[]{\label{fig3} The same model for the 1981 data set presented
in Fig.~\ref{fig2}b, in units of $\nu I_\nu$ to show the distribution
of power across the frequency bands.  Above $10^{18}$ Hz one can see
an analogy to the ``camel back'' feature known from blazar spectra.}
\end{figure}

\subsection{1981 data set, with acceleration}

Therefore we want to focus on solutions where the particles in the jet are
accelerated into a power-law distribution, as is typical in AGN.  This
solves the energy budget problem for the jet power, although we still
need a very luminous thermal contribution.  

We show four representative fits in Fig.~\ref{fig2}, all with a nozzle
radius of $3 r_g$, which is more appropriate for the assumption that
the jet originates at the edge of an accretion flow.  As discussed
above, we fix $L_{\rm d}=0.1 L_{\rm Edd}$ with $T_{\rm in}=10^6$ and
$r_{\rm out}=10^5 r_g$ and then fit the optical data with the
irradiated component, leaving the intercepted fraction of $Q_{\rm j}$,
$L_{\rm BB}$, as a free parameter in the fit.

In panel (a) of Fig.~\ref{fig2}, we show the resulting spectrum from a
jet with mildly relativistic electrons, similar to the model shown in
Fig~\ref{fig1}a.  The short-dashed line shows their contribution
before encountering the acceleration region.  Compared to
Fig.~\ref{fig1} the nozzle ``hump'' lies at a higher frequency due to
the more compact nozzle scale, and the slope is more inverted
primarily to the greatly increased angle to the line-of-sight
necessitated by the radio/IR fit, $\theta_i=55^\circ$.  The slope is
flatter in the post-accelerated component (shown with the
dot-dashed line), because of the effects of reacceleration and
particle losses.  Synchrotron from the accelerated particles dominates
the entire spectrum with the exception of the optical, with the IR and
X-rays stemming from the region near the start of the acceleration
$z_{\rm sh}$, while the radio comes from much further out along the
jet.  For a fixed $T_{\rm e}$, $Q_{\rm j}$ and $\theta_i$ are found by
fitting the radio-IR spectrum, which then uniquely determines $z_{\rm
sh}$ by fitting the IR turnover and extension to the X-rays.  This
also determines the slope of the accelerated particles, $p$.  The jet
parameter values for this mildly relativistic case are: $T_{\rm
e}=2\times10^9 K$, $Q_{\rm j}=0.62 L_{\rm Edd}$, $\theta_i=55^\circ$,
$z_{\rm sh}=2.15\times10^3 r_g$ and $p=2.15$.  The IC emission is
dominated by upscattered thermal disk photons and is well under the
synchrotron contribution.

The problem with this model is that---as in the model from
Fig.~\ref{fig1}a---the required power in the jet $Q_{\rm j}$ is
unrealistically large, in order to compensate for the low electron
temperature.  This imbalance also leads to the magnetic field energy
density in the jet being a few orders of magnitude super-equipartition
with the hot, radiating electrons.  While many jet formation models
assume the magnetic energy density is in equipartition with the
kinetic energy near the base of the jet (e.g.,
\citealt{BlandfordPayne1982,LiChiuehBegelman1992}), which for a maximal
jet would lead to magnetic domination, this may be rather
extreme.

In Fig.~\ref{fig2}b, we consider the highest temperature electrons
allowed by this model, above which the X-ray predictions (from either
synchrotron or IC from the jet) are too high, $T_{\rm e}=7\times10^9$
K. The jet free parameters are now: $Q_{\rm j}=0.27
L_{\rm Edd}$, $\theta_i=55^\circ$, $z_{\rm sh}=1.75\times10^3 r_g$ and
$p=2.15$.  Raising the temperature thus serves mainly to reduce the
power requirements and changes the location of the shock region
slightly.  With the same SD parameters as in (a), the irradiation
component now comprises $0.4 Q_{\rm j}$, still rather large but again
given the uncertainties in the mass and outer radius, more viable.
The hotter electrons now also contribute to a greater degree via IC
upscattering, contributing some of the soft X-ray emission.

While there is no measurement of the inclination angle, some papers
favor a lower value than what we find for our best fit above (e.g.,
\citealt{Cowleyetal2002}).  In Fig.~\ref{fig2}c, we show what our model
predicts for $\theta_i=15^\circ$ and the same temperature as in panel
(b).  The decreased angle greatly reduces the power requirements in
the jet, but at the same time flattens the predicted radio-IR slope
significantly, making a good radio/IR fit impossible.  The parameters
are $Q_{\rm j}=3.2\times10^{-2} L_{\rm Edd}$, $z_{\rm
sh}=2.15\times10^3 r_g$ and $p=2.15$.  Although this radio point is
only from an extrapolation of the radio/X-ray correlation relation
(C00; C02), the slope is remarkably well-defined at these higher
luminosities.  On the other hand, $\theta_i=55^\circ$ does give good
fits to most of the other data sets discussed below.  Because the
beaming reduces the power requirements for the jet, a low inclination
angle also means that more power must irradiate the outer disk than is
available, making this model problematic.

Finally in Fig.~\ref{fig2}d we consider the case of a power-law of
particles entering the base of the jet from the accretion flow instead
of thermal particles, as may occur if acceleration also takes place at an
accretion shock.  Instead of defining a temperature, we take the
minimum electron Lorentz factor $\gamma_{\rm e, min}=1$, leaving
$\gamma_{\rm e, max}$ a free parameter.  In order to prevent the
nozzle emission from violating the constraints of the soft X-ray data,
we find $\gamma_{\rm e, max} \sim 10$, which would require extreme
radiative cooling to truncate the accelerated power-law for this
scenario to be realistic.  This type of scenario may be necessary for
certain low-luminosity AGN (see, e.g., \citealt{Yuanetal2002}). The
other jet parameters are $Q_{\rm j}=0.55 L_{\rm Edd}$, $z_{\rm
sh}=2.15\times10^3 r_g$, $\theta_i=55^\circ$ and $p=2.15$.  The high
required jet power means that the irradiated component needs only
$0.02 Q_{\rm j}$ to account for the optical data.

In this last model, the IC emission still lies under the synchrotron
component, but not by much, and has the correct slope.  In a different
geometry where the SD underlies the corona/jet base, this may
provide the best alternative for a solution where the IC dominates the
synchrotron emission in the X-rays.

In all four model runs shown here, we needed a rather large fraction
of the jet power to be intercepted by the outer disk and reradiated.
If we allowed the SD to radiate $0.2 L_{\rm Edd}$ away, we could lower
the necessary $L_{\rm BB}$, but then again we would be basically at
the Eddington accretion rate, as discussed for the non-acceleration
case.  This problem is clearly not very constrained for this source,
and a central mass of $\sim15$ instead of 5 $M_\odot$ (as in, e.g.,
GRS~1915+105; \citealt{Greineretal2001}), would easily resolve these issues.

In Fig.~\ref{fig3} we show the same model as in Fig.~\ref{fig2}b with
the flux multiplied by the frequency to show the distribution of power
across the wavebands.  One can see that most power seems to fall in
the disk component, with the synchrotron then creating the second peak
around the cutoff.  The IC component is in fact dominated by the
upscattered disk photons, but this is hidden beneath the synchrotron
emission.  The self-Comptonized synchrotron component peaks at
$10^{23}$ Hz, or $\sim 400$ MeV, corresponding to the synchrotron peak
upscattered by the hot electrons near the shock.  At high frequencies,
the shape is similar to the ``camel back'' spectra seen in blazars
(e.g., \citealt{Fossatietal1998}), and in fact at this highest observed
flux level is above the sensitivity of GLAST for a one year all-sky
survey at 100 MeV.  The less powerful epochs of this source presented
in the next section may not be so observable, however, as they are
orders of magnitude lower.

In conclusion, we have shown that in order to explain the
low-frequency data from the 1981 observation of \gx, a jet solution
must include some form of particle acceleration.  And further, once
that condition is fulfilled, X-rays from jet synchrotron (in addition
to IC) are unavoidable as long as IC emission is not the main cooling
channel.  But under the assumption of a disk geometry where a SD
transforms to an optically thin, hot accretion flow at some inner
radius, synchrotron cooling will dominate in a maximal jet.  It is
possible, however, that in the alternative geometry of a corona/jet
base lying directly on top of the SD could lead to so much IC cooling
that the synchrotron cutoff would occur below $\sim 10^{17}$ Hz, thus
accommodating a dominant IC contribution in the X-rays.  This would
not, however, explain the interesting coincidence of IR turnover/X-ray
spectrum seen here and now also in several sources in addition to
XTE~J1118+480.  It may also be difficult to explain the radio/X-ray
correlation with an IC model (see Sect.~\ref{corsec}).  We will,
however, soon explore combined jet/corona solutions using lower
temperature cyclo-synchrotron and IC scattering in different
geometries.

We have explored the relativistic electron parameter range for the
1981 data set and find that the most reasonable parameters are
obtained for the case of a fully relativistic initial thermal electron
distribution.  Besides requiring too much power, a mildly relativistic
initial electron distribution would also result in the magnetic energy
density being much higher (and further out of equipartition with) the
accelerated particles at the shock, more so than would likely be
expected for magnetic confinement.  For these reasons, in this paper
we consider only the higher temperature solutions for all fits,
corresponding to the type of fit shown in Fig.~\ref{fig2}b.

\subsection{Other data sets}

For the 12 data sets compiled during the other LHS episodes of \gx, we
can either optimize each fit individually leaving all parameters free,
or fix as many parameters as possible to see if essentially the
same model can account for all observed spectra.  We here choose to
take this latter approach, in order to explore the fundamental physics
of the model.  We will fix some of the parameters to those best
constrained by the 1981 set and then study the affect of changing only
two jet parameters.  Our emphasis is on understanding the radio/X-ray
correlations, and thus the jet component.  However, we address the SD
contribution as well, and attempt to test the jet/disk connection by
finding a scaling solution.  We set $L_{\rm BB}=0.1 Q_{\rm j}$ for all
fits, and then fit $L_{\rm d}$ to the optical data, when present.

While it seems clear that there are small variations in the soft X-ray
spectra of these data sets (this is discussed in NWD, who fit a broken
power law---we are fitting the lower energy segment), we choose to fix
the accelerated spectral index to $p=2.15$ as in 1981.  Similarly,
while there are obvious changes in the SD contribution, they seem
consistent with the likely underlying accretion changes that also
affect the jet since both components decrease in luminosity with
increasing time.  We keep $T_{\rm in}$ fixed at $10^6$ K and $T_{\rm
e}$ fixed at $7\times10^9$ K based on the 1981 set, as a starting
point.  We will assume that the orientation of the jet remains fixed
at $55^\circ$ (statistically $57^\circ$ is the most likely angle) and
that the base of the jet is fixed at $r_0=3 r_{\rm g}$.  Thus the only
jet parameters allowed to vary are $Q_{\rm j}$ and $z_{\rm sh}$, and
the disk parameter $L_{\rm d}$ is then adjusted to fit the thermal
data, when present.

\begin{figure*}[h]
\centerline{\psfig{figure=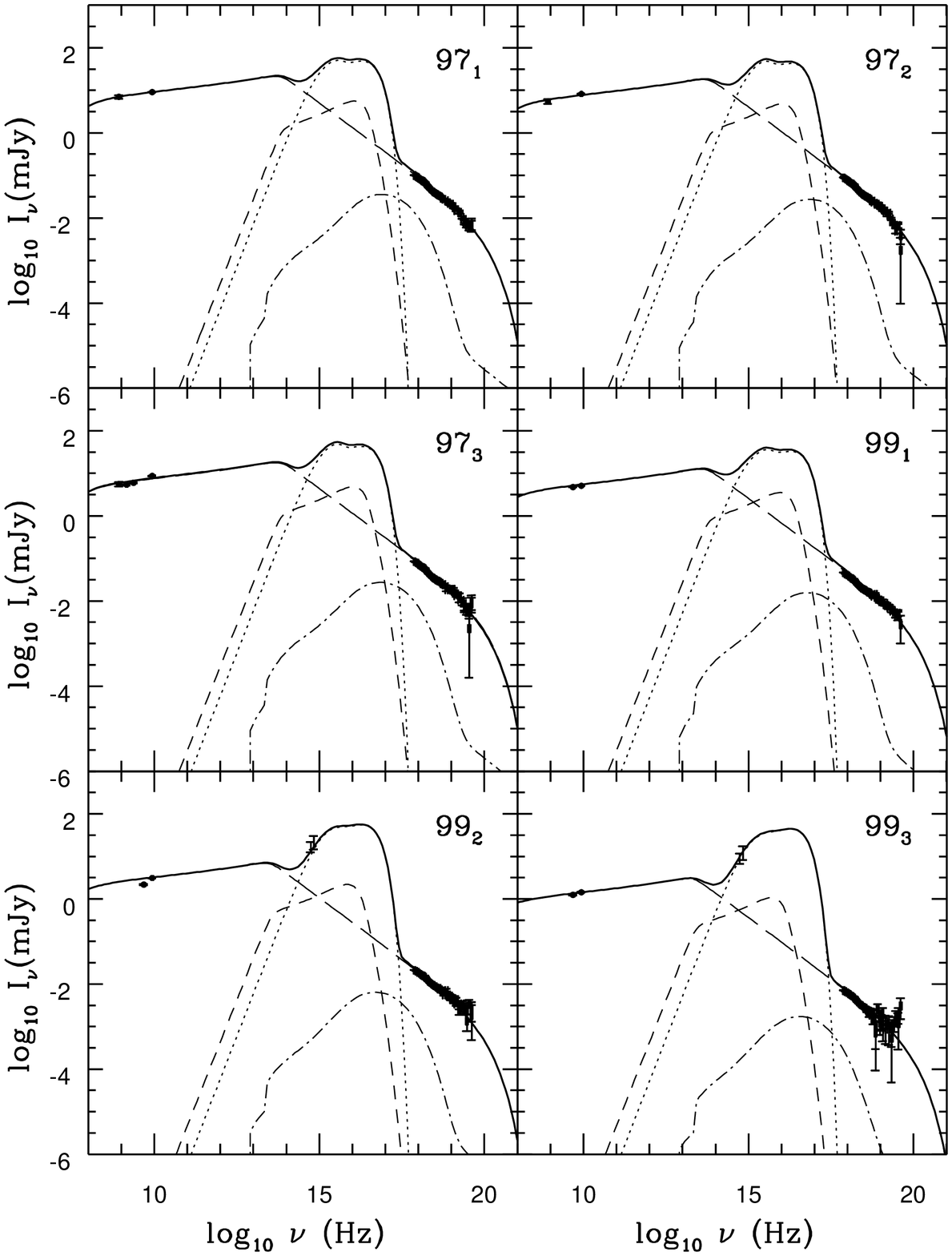,width=6in}}
\caption[]{\label{fig4a}(a) Composite figure with 6 data sets with labels
corresponding to the data sets of Table~\ref{tab1}, taken from W99,
C00, C02 and NWD.  The solid thicker line is the total spectrum, the
dotted line is the multi-temperature blackbody representing the outer
thermal disk plus a single blackbody for irradiation, the short dashed
line is the synchrotron emission from the jet before encountering the
shock acceleration region, and the long-dashed line is that from
after, and the dot-dashed line is the IC jet and disk emission
upscattered by the jet plasma.  Fit parameters are shown in
Table \ref{tab2}. }
\end{figure*}

\setcounter{figure}{3}
\begin{figure*}[h]
\centerline{\psfig{figure=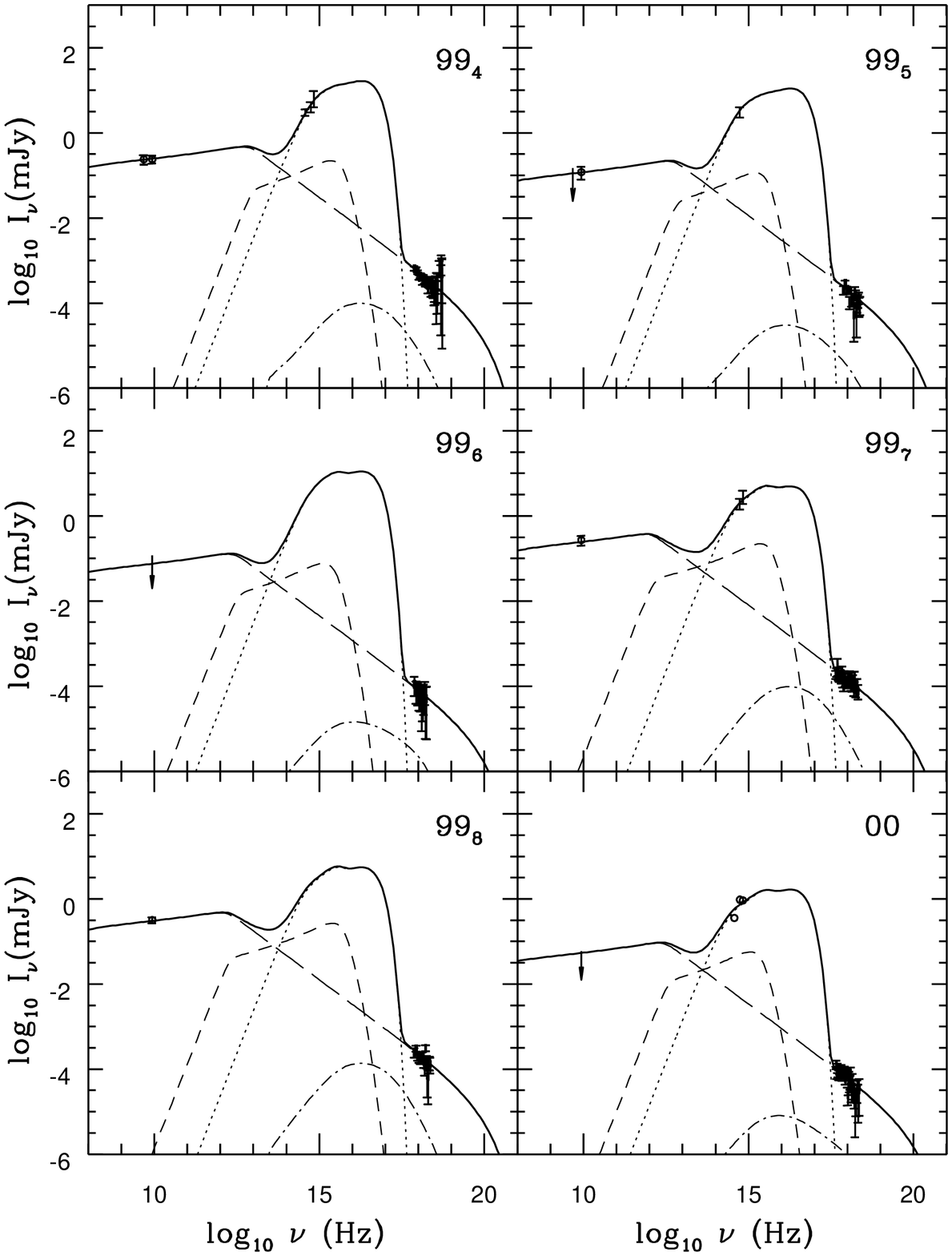,width=6in}}
\caption[]{\label{fig4b}(b) Composite figure with 6 data sets with labels
corresponding to the data sets of Table~\ref{tab1}, taken from
W99, C00, C02 and NWD.  The arrows represent $3-\sigma$ detection upper
limits.  The lines are the same as in Fig.~\ref{fig4a}a.  For the 'off'
state in 2000, the optical data are from three months prior to the X-ray
and radio observations, and should be seen as upper limits. }
\end{figure*}

Figs.~\ref{fig4a}a \& \ref{fig4b}b show the broadband spectral fits to
the data, with the corresponding fit parameters given in
Table~\ref{tab2}.  For the data sets without optical data constraining
the disk, $L_{\rm d}$ is rather arbitrary, and we scaled the
luminosity to the jet power.

Although the fits are surprisingly good given the limited fitting
parameters, there are obvious features in the X-ray waveband such as a
flattening at higher frequencies and peaks which our model cannot
account for.  These features are likely due to the impacting X-rays on
the cooler material of the SD, but we have not yet incorporated this
into our model.  The softest X-rays should not be affected by
these additional components, however, so we focus our fitting at the
lower X-ray frequencies.  For the lowest flux level data sets of 1999,
which we label $99_4$---$99_8$, as well as 2000, the error bars are
large enough to make the fitting more difficult.

\begin{figure}[t]
\centerline{\psfig{figure=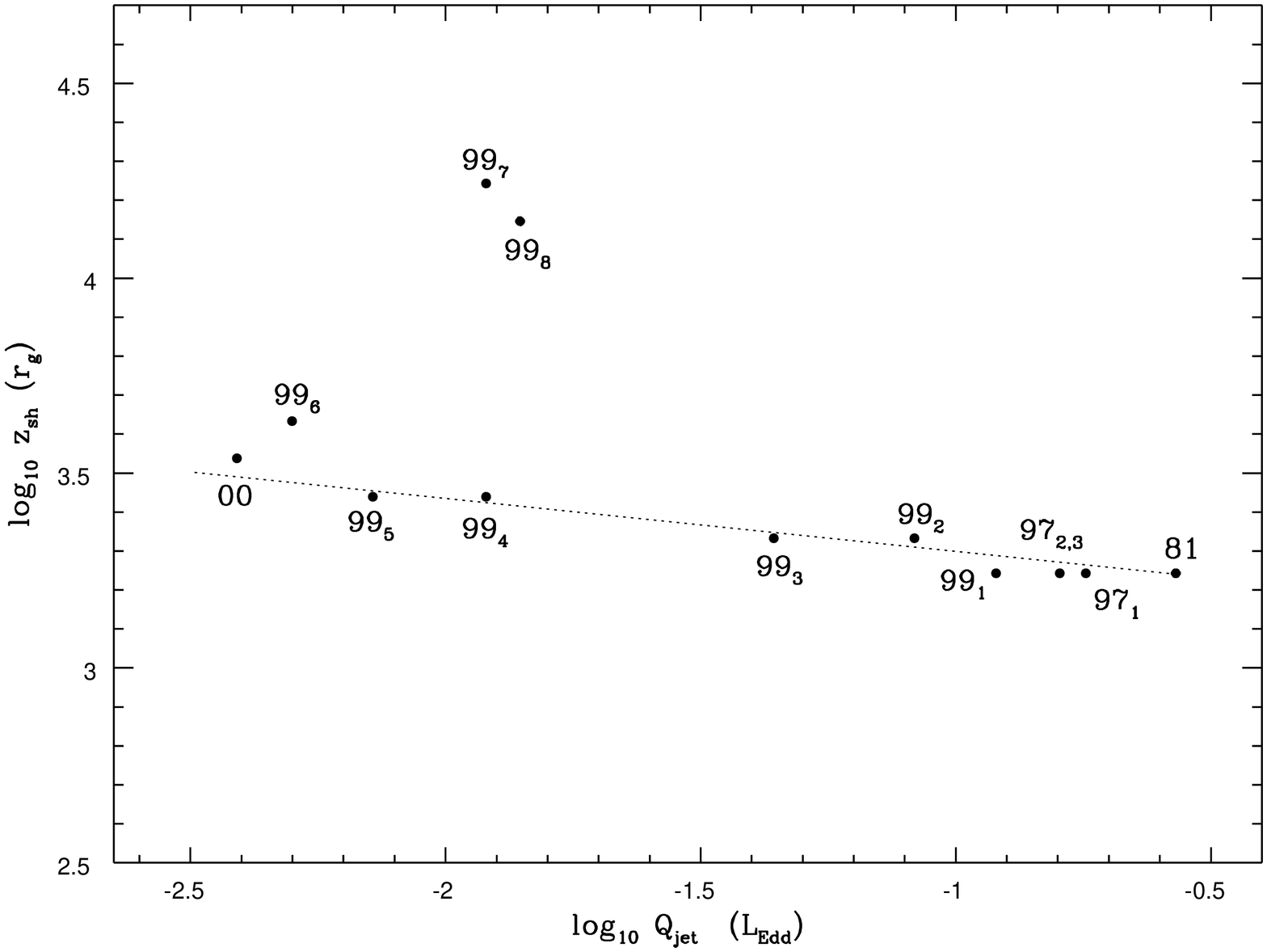,width=\figwidth,angle=-90}}
\caption[]{\label{fig5} Correlation between the input jet power
$Q_{\rm j}$ and the location of the first acceleration zone $z_{\rm
sh}$ for the fits in Figs. \ref{fig2} and \ref{fig4a}.  The dashed
line is to guide the eye for the upward trend in the data sets.  The
1981 point represents the fit shown in Fig.~\ref{fig2}b.  The $99_8$
data set is not strictly simultaneous, and the radio data are
simultaneous with a small reflare in the hard X-rays, which may
explain its excess.  $99_7$ may also be affected by this flare.}
\end{figure}

The most important feature to note is that the radio/X-ray
correlations detected in this source (C00; C02) can be explained by
the variation of only one main parameter in the jet model: the input
power.  If we plot $z_{\rm sh}$ against the jet power $Q_{\rm j}$
(Fig.~\ref{fig5}), we see that there is a slight dependence ($z_{\rm
sh} \propto Q_{\rm j}^{-0.135}$), but this likely results from our
rather artificially freezing of the other parameters.  This small
dependence is rather amazing when one thinks that $Q_{\rm j}$ is
changing by about three orders of magnitude over the same range, if
1981 is included.

\begin{table*}[t]
\caption{\label{tab2}Fit Parameters for Figures 3a and 3b}
\begin{tabular}{llcl}
\hline
\hline
{\bf Data Set} & {\bf $Q_{\rm j}$}  & {\bf
$z_{\rm sh}$} & {\bf $L_{\rm d}$} \\
& ($L_{\rm Edd}$) & ($r_g$) & ($L_{\rm Edd}$) \\
\hline
$97_1$* & 0.18               & $1.8\times10^3$  & $8.8\times10^{-2}$ \\
$97_2$* & 0.16               & $1.8\times10^3$  & $7.8\times10^{-2}$ \\
$97_3$* & 0.16               & $1.8\times10^3$  & $7.8\times10^{-2}$ \\
$99_1$* & 0.12               & $1.8\times10^3$  & $6.0\times10^{-2}$ \\
$99_2$  & $8.3\times10^{-2}$ & $2.2\times10^3$  & $1.0\times10^{-1}$ \\
$99_3$  & $4.4\times10^{-2}$ & $2.2\times10^3$  &  $8.0\times10^{-2}$ \\
$99_4$  & $1.2\times10^{-2}$ & $2.8\times10^3$  & $3.0\times10^{-2}$ \\
$99_5$  & $7.2\times10^{-3}$ & $2.8\times10^3$  & $2.0\times10^{-2}$ \\
$99_6$* & $5.0\times10^{-3}$ & $4.3\times10^3$  & $2.0\times10^{-2}$ \\
$99_7$  & $1.2\times10^{-2}$ & $1.8\times10^4$   & $9.0\times10^{-3}$ \\
$99_8$* & $1.4\times10^{-2}$ & $1.4\times10^4$  & $1.0\times10^{-2}$ \\
$00^a$  & $3.9\times10^{-3}$ & $3.5\times10^3$  & $3.0\times 10^{-3}$ \\
\hline
\end{tabular}

\small *Data sets where there is no direct constraint on the thermal disk
component.\\
$^a$The optical data constraining the SD are not simultaneous.
\end{table*}

\section{Understanding the radio/X-ray correlations}\label{corsec}

\begin{figure}[t]
\centerline{\psfig{figure=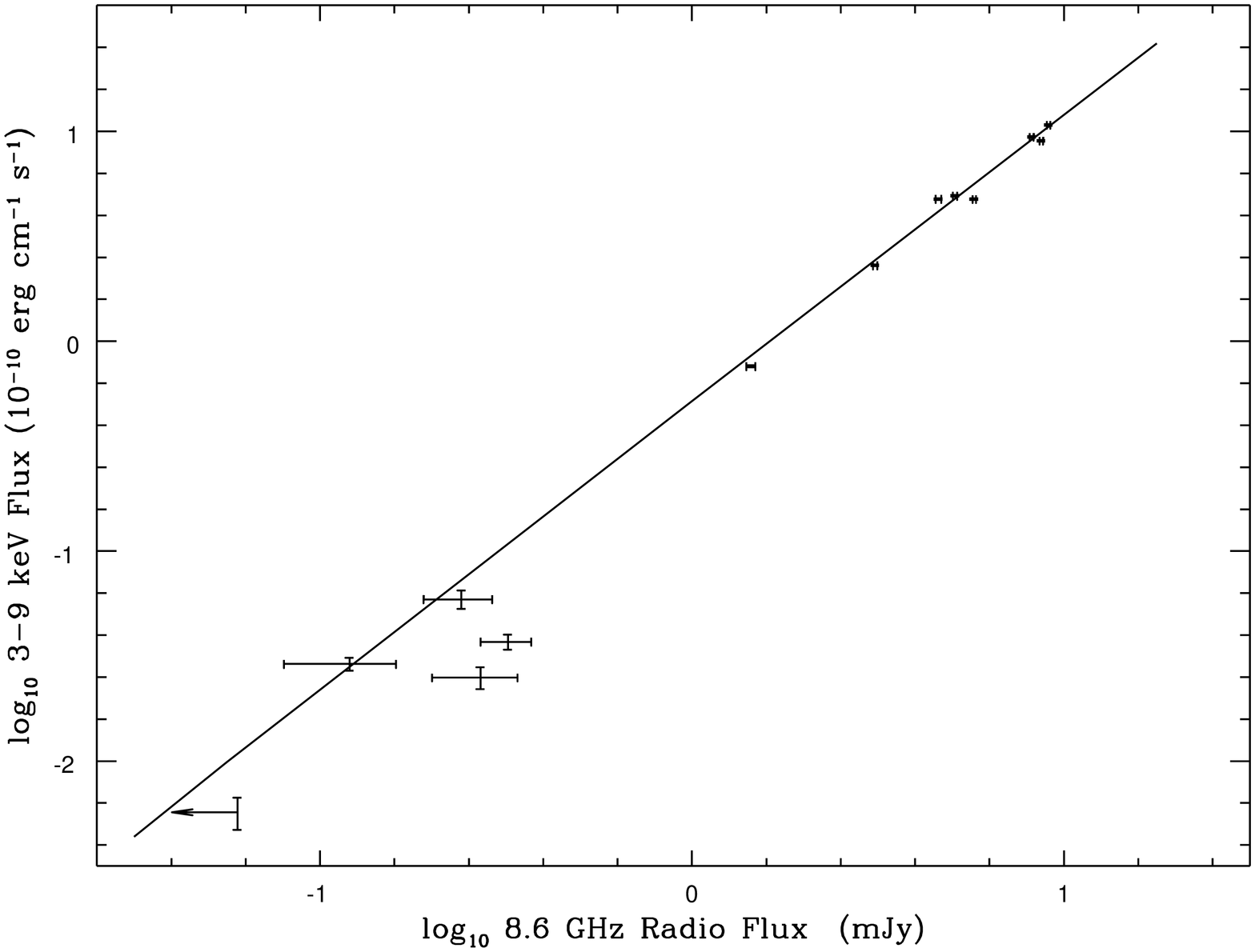,width=\figwidth,angle=-90}}
\caption[]{\label{fig6} The model-predicted radio (8.6 GHz)/X-ray
(integrated 3-9 keV emission) correlation (solid line).  The data are
from C00 \& C02.  For this figure, $z_{\rm sh}$ is fixed at $\sim
1.8\times10^3 r_g$ to match the 1981 highest flux result, and then the
only changing parameter is the power input into the jet $Q_{\rm j}$.
The arrow represents $3-\sigma$ upper limits.}
\end{figure}

It seems likely that the real source of the variations is the
power fed into the jet.  In Fig.~\ref{fig6} we plot the prediction of
our model against the data for the correlation between the 8.6 GHz
radio flux and the integrated 3-9 keV X-rays (C02).  We fix $z_{\rm
sh}\sim1.8\times10^3 r_g$, to match the highest flux data point of
1981, which is not shown on this figure because the radio was
not simultaneously measured.  Then we adjust only $Q_{\rm j}$ while
holding {\em every} other parameter constant, in order to fit the
radio flux, and plot the resulting integrated 3-9 keV X-ray flux. 
In removing the $z_{\rm sh} \propto Q_{\rm j}^{-0.135}$ dependence
mentioned above, we are likely compromising the quality of the fit to
some degree, but this could be compensated for if we allowed
particularly the spectral index to vary.

By just changing the jet power we obtain a very good fit to the
correlation data, with the exception of the two data sets $99_7$ and
$99_8$.  In fact, the radio data in $99_8$ was taken a few days after
the X-rays, and coincided with a small reflare in the hard X-rays (see
Fig.~15 in C00).  This is the likely explanation of the excess radio
flux.  Similarly, the reflare may also have an affect on $99_7$, if
there is, e.g., a lag between the changing fluxes.

The results shown in Fig.~\ref{fig6} provide strong support for a jet
synchrotron model, since it correctly predicts the radio/X-ray
correlation over orders of magnitude in flux levels, almost exactly
matching the data.  The fit is better at the higher flux end, where
the error bars are smaller.  The exact dependence (excluding $99_7$ and
$99_8$) goes as $F_{\rm X} \propto F_{\rm R}^{m}$, where
$m=1.41$, $F_{\rm R}$ is the 8.6 GHz radio flux and $F_{\rm X}$ is the
integrated 3-9 keV X-ray flux.

The value for $m$ in fact follows directly from analytic
predictions of the jet model.  Eq. (52) in \citet{FalckeBiermann1995}
shows\footnote{In \citet{FalckeBiermann1995}, the quantities which we
label $\nu_{\rm SSA}$, $Q_{\rm j}$ and $F_{\rm SSA}$ correspond to
$\nu_{\rm s,obs}$, $q_{\rm j/1}L_{\rm 46}$ and $L_{\nu_s, {\rm obs}}$,
respectively.} that if all parameters except the power are fixed, as
we have done to produce Fig.~\ref{fig6}, the frequency where the
optically thick flat-to-inverted spectrum turns over to the optically
thin regime depends on the power as
\begin{equation}\label{nussa}
\nu_{\rm SSA} \propto Q_{\rm j}^{2/3}. 
\end{equation}
Similarly Eq. (56) (ibid.)  shows that under the same
freezing of other parameters, the flux at the turnover depends on the
power as
\begin{equation}\label{fssa}
F_{\rm SSA} \propto Q_{\rm j}^{17/12}.
\end{equation} 
These relations can then be normalized using the 1981 data set, where
the flux and turnover frequency are visible.

Knowing these relations, and the spectral indices of the optically
thick and thin components from the data, one can roughly calculate the
expected correlation slope.  Because the spectral indices for the
radio-IR, $\alpha_{\rm RIR}$, and X-ray,  $\alpha_{\rm X}$, do vary
slightly over \gx's observation history, taking constant values will
make this less exact. We now want to calculate the slope between two points
along the correlation curve, analogous to the model plotted in
Fig.~\ref{fig6}.  Each point represents a different value of $Q_{\rm
jet}$, corresponding to different values of $\nu_{\rm SSA}$ and
$F_{\rm SSA}$.  The 8.6 GHz
radio flux is then
\begin{equation}
 F_{8.6}=F_{\rm SSA} \left(\frac{\nu_{\rm SSA}}{8.6 \;{\rm
GHz}}\right)^{\alpha_{\rm RIR}},
\end{equation} 
and then using Eqs. \ref{nussa} \& \ref{fssa}:
\begin{equation}
\log_{10} F_{8.6} = \log_{10}C_1 +  \frac{17}{12} \log_{10} Q_{\rm j} -\frac{2}{3}\alpha_{\rm RIR}
\log_{10} Q_{\rm j},
\end{equation} 
where $C_1$ absorbs the exact dependences of the flux and frequency at
the turnover on $Q_{\rm j}$, and other constants.

Similarly, the X-ray flux at any frequency (as long as it falls on the
power-law) is
\begin{equation} 
F_{\rm X} = F_{\rm SSA} \left(\frac{\nu_{\rm
SSA}}{\nu_{\rm X}}\right)^{\alpha_{\rm X}},
\end{equation} 
which after using the forms from Eqs. (\ref{nussa}) \& (\ref{fssa})
and integrating from 3-9 keV becomes
\begin{multline} 
\log_{10} \int F_{\rm X} d\nu =
\log_{10}C_2 +\frac{17}{12}\log_{10} Q_{\rm j} -\\ \frac{2}{3}\alpha_{\rm X}
\log_{10} Q_{\rm j},
\end{multline}
where again $C_2$ includes the constants of integration and the
dependencies on $Q_{\rm j}$.  

For any two values of the jet power, $Q_A$ and $Q_B$, the correlation
slope is then 
\begin{eqnarray}
m & = & \frac{\log_{10} \int F_{\rm X, B} - \log_{10}
\int F_{\rm X, A}}{\log_{10} F_{\rm 8.6, B}-\log_{10} F_{\rm 8.6,
A}}\\
& = & \frac{\frac{17}{12}-\frac{2}{3}\alpha_{\rm
X}}{\frac{17}{12}-\frac{2}{3}\alpha_{\rm RIR}},
\end{eqnarray}
after algebraic cancellations.  Taking the values from the power-law
fits to the 1981 data set from CF02, $\alpha_{\rm RIR}\sim0.15$ and
$\alpha_{\rm X}\sim-0.6$, we find $\alpha_{\rm cor}=1.38$, which
considering the simplifications in the model and our ignoring of the
spectral variations, is surprisingly close to the value of 1.4 found
for \gx.  In addition, this type of radio/X-ray correlation has also
been seen in V404 Cygni, with exactly the same slope
(\citealt{GalloFenderPooley2002}; Gallo et al., MNRAS, in prep.).  Our preliminary models of this source also
show a turnover in the IR band and have similar jet parameters
(Markoff et al. 2002, in prep.).

The fact that the well-determined slope $m$ extrapolates to the lower
flux data also gives us greater confidence in the subtraction of the
putative ``background source'' discussed in Sect.~2.  Note also that
the lowest flux point is from BeppoSAX, and therefore does \emph{not}
rely on the subtraction of a background source.

It is possible that an IC corona model could also explain the correlation
slope, if the radio flux scales with the power (assumed to be linearly
related to $\dot M$) as expected from the jet and the X-ray flux
scales with ${\dot M}^2$.  This may be the case in some optically thin
accretion solutions.  However, it has yet to be shown that a self
consistent solution can be found which also well fits all the
broadband data available for this source.  Similarly this would not
explain why the X-rays trace back to the turnover in the IR as  1981 data, or why most if not all LHS sources show this turnover
coincidence.

\section{Conclusions}

We show that a model comprising a dominantly synchrotron jet
component, in combination with an optically thin accretion flow
transitioning to a standard thin disk, is able to explain the
broadband spectral data from 13 observations of \gx.  At the same
time, the model can easily explain the $m \sim 1.4$ slope of the
radio/X-ray correlations by changing only the power input into the
jet.  This input power, $Q_{\rm j}$, is assumed to be proportional to
the accretion rate, and varies from $L_{\rm Edd}\sim 0.003-0.3$ for
a 5$M_\odot$ black hole.  Above the highest luminosity, the source
would likely transition to the HSS.

This model assumes an underlying disk geometry in which the standard
thin disk exists beyond $\sim 100-1000\; r_g$, and has a relatively low
accretion rate.  In a situation where the SD extends all the way down
to the last stable orbit, the photon field at the base of the jet may
be high enough to affect our conclusions, which are dependent on
synchrotron losses dominating the cooling.  If the available photon
field for inverse Compton upscattering becomes large enough, it may
truncate the accelerated particle distribution and lessen the extent
of the synchrotron emission.  

With the exception of this model, the non-thermal X-ray spectral
component in all XRB states has been modeled in terms of inverse
Compton processes in a hot corona, located either above or within the
standard thin disk.  These classes of models have significant success
explaining both the spectral features and timing characteristics in
the X-ray waveband, but fail to address the radio/IR spectra
which in at least \gx, V404 Cyg and Cyg X-1 seem intimately linked to
the higher frequencies.

However, the existence of a hot, magnetized plasma at the base of the
jet leads to obvious considerations of a unification scheme between it
and this concept of a corona, which until now has not been explicitly
observed.  Both scenarios (jet and corona) address important features
from the observations, but either cannot or have not yet attempted to
account for everything.  We feel that a concatenation of these two
components would be fruitful as the next stage of our investigation.
It is, however, important to note that under reasonable physical
assumptions, the jet can easily produce X-ray emission via synchrotron
radiation, and has significant success explaining the data when it
does.  This possibility must therefore be considered in spectral
modeling; whether it really dominates the entire spectrum is a
question of the environment and local acceleration conditions.  It may
be that it indeed dominates only in those sources which show only weak
reflection features (e.g., XTE~J1118+480; \citealt{Milleretal2001}),
but it likely has a wider impact in all sources than has so far been
appreciated.

Interestingly, the parameters we derive for the power and location of
the shock are roughly consistent with earlier results modeling
XTE~J1118+480, which because of the higher temperature
($T_e=2\times10^{10}$ K) resulted in a lower necessary power input
into the jet.  In this source we found $z_{\rm sh}\sim10^2 r_g$ as
compared to the $\sim10^3 r_g$ found here, however considering that
the jet can extend beyond $10^{10} r_g$, this range is quite small.
In our recent modeling of other LHS BHC sources, we are in fact
finding that all sources with simultaneous radio/X-ray data seem to
require acceleration to begin in this same range, and we will discuss
the physical implications of this elsewhere.  For these sources we
also find, as in \gx and XTE~J1118+480, that if the X-rays are traced
back to lower frequencies, the optically thick-to-thin turnover always
occurs in the IR range.  If this turnover coincidence is real, it
offers us the chance to explore the physical conditions at the
acceleration zone of XRB jets.

This work illustrates how critical simultaneous multiwavelength
observations are to increasing our understanding of the physics in
these sources.  With the advent of higher energy missions such as {\em
GLAST}, {\em INTEGRAL} and {\em ASTRO E-2}, soon we can hopefully begin
to probe the contribution of these smaller jets to the hard
X-ray and $\gamma$-ray bands.

\begin{acknowledgements}
This work is partially supported by AUGER Theory Grant O5CU1ERA/3 from
the BMBF (S.M.).  S.M. would like to thank Tom Maccarone and David
Meier for very helpful discussions.
\end{acknowledgements}

\bibliography{aamnemonic,h3854} \bibliographystyle{aa}

\end{document}